\documentclass[11pt]{article}
\usepackage{orcidlink}
\usepackage[margin=1in]{geometry}
\usepackage{enumitem}
\usepackage[sort,compress]{cite}
\usepackage{xcolor}
\usepackage{hyperref}
\hypersetup{colorlinks=false,allbordercolors=blue,pdfborderstyle={/S/U/W 1}}
\usepackage{amsmath,amssymb}
\usepackage{physics}
\usepackage{graphicx}
\usepackage{tikz}
\usetikzlibrary{quantikz}
\usetikzlibrary{calc, angles, quotes}
\usepackage{booktabs}
\usepackage{url}
\usepackage{multirow}

\newenvironment{itemlist}{\begin{itemize}}{\end{itemize}}
\newenvironment{romanlist}[1][(i)]{\begin{enumerate}[label=(\roman*),start=2]}{\end{enumerate}}
\newcommand{\keywords}[1]{\par\medskip\noindent\textbf{Keywords:} #1\par\medskip}
\newcommand{\tbl}[2]{\caption{#1}#2}
\newcommand{\colrule}{\midrule}
\newcommand{\botrule}{\bottomrule}

\begin{document}

\title{A Dual-Uplink Dual-Satellite Architecture \\ for Entanglement Distribution}

\author{Charilaos C. Angelidis\,\orcidlink{0009-0006-9564-6518}$^{1}$
and Ioannis K. Marmorkos$^{2}$
\thanks{Email correspondence: cangel03\@ucy.ac.cy and imarm\@ihu.gr}\\[0.35em]\small $^{1}$ Department of Electrical and Computer Engineering, University of Cyprus, Nicosia, Cyprus\\\small $^{2}$ Department of Information and Electronic Engineering, International Hellenic University, Thessaloniki, Greece}

\date{}
\maketitle

\begin{abstract}

Satellite-based quantum links are a promising enabler of long-distance entanglement distribution, overcoming the exponential losses of optical fibers. In uplink architectures, entangled photons are generated at ground stations and transmitted to orbiting satellites, reducing space-segment complexity but introducing significant atmospheric and geometric losses. Single-satellite dual-uplink configurations are additionally limited by the requirement of simultaneous visibility between both ground stations and the satellite.

In this work, we investigate a dual-uplink dual-satellite network architecture and compare its performance with the single-satellite scheme under identical physical assumptions. The proposed topology introduces an inter-satellite link (ISL) that decouples the ground-station separation from the atmospheric uplink geometry. A realistic channel model is employed, incorporating geometric attenuation, atmospheric turbulence, background noise, and Bell-state measurement efficiency. The overall success probability and practical fidelity are evaluated as functions of satellite altitude and ground-station separation.

Our results show that while the single-satellite configuration is suitable for regional-scale networks, it becomes infeasible for large separations due to geometric constraints. The proposed dual-satellite architecture mitigates this limitation and significantly extends the achievable communication distance, representing a key step toward scalable quantum internet infrastructures.

\end{abstract}

\keywords{Quantum network architecture; Satellite quantum communications; Inter-satellite links; Entanglement distribution; Network scalability; Performance evaluation.}

\section{Introduction} \label{sec1}

Quantum optical fiber networks suffer from fundamental distance limitations due to the exponential attenuation of photons \cite{ref1,ref2}. Satellite-based quantum networks have therefore emerged as a promising alternative for enabling long-distance quantum communications \cite{ref3,ref4,ref5,ref6}. In the context of the quantum internet, quantum entanglement constitutes a key resource \cite{ref7,ref8,ref9,ref10,ref11,ref12,ref13,ref14}, and scalable satellite quantum networks are essential for its reliable distribution over global distances.

Most existing studies have focused on downlink architectures \cite{ref15,ref16,ref17,ref18}, where entangled photon pairs are generated onboard a satellite and transmitted to distant ground stations. In contrast, uplink architectures, in which entanglement is generated at the ground stations and transmitted toward a satellite, experience higher losses, since the most turbulent and noisy segment of the optical channel is encountered at the beginning of the signal propagation \cite{Pirandola2020SatelliteQC,Bonato2009FeasibilityOS,ref21}. 

Despite these drawbacks, uplink architectures offer significant practical advantages, as the stringent power, mass, and complexity constraints of satellites represent major challenges for large-scale quantum network deployment. Ground stations, on the other hand, can support substantially higher power budgets and more sophisticated optical systems, enabling enhanced control over entanglement generation while simultaneously reducing satellite payload complexity \cite{ref22}.

Recent work \cite{Srikara2024QuantumED} has investigated a dual-uplink single-satellite architecture for entanglement distribution under realistic physical conditions, demonstrating that such a configuration can remain viable despite the increased loss profile associated with uplink transmission. However, single-satellite architectures are fundamentally limited by the geometric requirement of simultaneous visibility. As the distance between ground stations increases, the satellite must reach higher altitudes to maintain line-of-sight, which drastically increases path loss and reduces the entanglement generation rate to impractical levels.

In this work, we propose a Dual-Uplink Dual-Satellite architecture to relax these geometric constraints. The main contributions of this paper can be summarized as follows:

\begin{itemlist}
\item We propose a novel network topology that decouples the ground station separation distance geometrically from the atmospheric uplink path, allowing for near-zenith transmission regardless of the total communication range.
\item We develop a comprehensive end-to-end channel model that incorporates atmospheric turbulence, geometric attenuation, and, crucially, the deterministic diffraction losses of the Inter-Satellite Link (ISL).
\item We demonstrate via numerical simulations that the proposed architecture maintains operational fidelity for ground distances exceeding 2000 km, significantly outperforming the single-satellite baseline, which becomes infeasible beyond 1000 km.
\end{itemlist}

The remainder of this paper is organized as follows. Section \ref{sec:related_work} provides a brief overview of the related literature. Section \ref{sec2} details the physical channel model and measurement setup, including atmospheric and noise effects. Section \ref{sec4} introduces the proposed dual-uplink dual-satellite architecture and discusses the simulation parameters. Section \ref{sec5} presents the numerical results and a comparative analysis of the two architectures. Finally, Section \ref{sec6} concludes the paper.

\section{Literature Review}

\label{sec:related_work}

The feasibility of space-based quantum communication was fundamentally proven by the pioneering Micius satellite mission. Yin {\it et al.} \cite{ref15} successfully demonstrated satellite-based entanglement distribution to two ground stations separated by 1200 kilometers, while Liao {\it et al.} \cite{ref16} achieved satellite-to-ground quantum key distribution (QKD). These early architectures predominantly utilized a downlink approach, where complex quantum sources are housed onboard the satellite.

In recent years, there has been a growing research interest in exploring uplink architectures to simplify the space segment. For instance, studies such as the one by Neumann {\it et al.} \cite{ref22} have shown that utilizing simple, cost-effective satellites purely as measurement nodes for ground-based entanglement sources offers significant advantages in terms of maintenance and scalability. At the same time, researchers have highlighted the severe physical challenges associated with this approach. Specifically, Liorni {\it et al.} \cite{52} extensively analyzed the atmospheric effects on such links, demonstrating that uplink optical channels suffer from profound beam wandering and broadening, as the signal encounters the densest parts of the atmosphere immediately upon transmission.

Recently, Srikara {\it et al.} \cite{Srikara2024QuantumED} provided a comprehensive evaluation of quantum entanglement distribution via uplink satellite channels using a single-satellite architecture. Their study highlighted that while uplink entanglement is feasible, the system is heavily constrained by the geometric requirement that both ground stations must simultaneously maintain optimal line-of-sight with the single orbiting node. This rigid geometric dependence drastically reduces the effective coverage and limits the maximum achievable ground distance, before the link fidelity drops below practical thresholds.

Motivated by these limitations, this paper explores a dual-uplink, dual-satellite architecture for entanglement distribution. The first satellite simply reflects the photon it accepts from the first ground station $A$ to the second satellite. The second satellite performs a Bell-State Measurement once it receives a photon from the second ground station $B$. By introducing a second satellite and an inter-satellite optical link (ISL) operating in the vacuum of space, we decouple the ground stations from the strict single-node geometry. This approach effectively bypasses the severe atmospheric and horizon constraints of conventional models, paving the way for viable quantum entanglement distribution across much larger, intercontinental distances.

\section{Physical Channel Modeling and Measurement Setup}\label{sec2}

\subsection{Link Geometry}

The geometric configuration of the ground-to-space and space-to-space links is a fundamental determinant of signal attenuation, as it directly governs both the free-space propagation distance and the effective atmospheric path length. 

\begin{figure}[t]

\centerline{
    \begin{tikzpicture}[scale=1.2, >=latex]
        
        \coordinate (O) at (0,0); 
        \def\Re{3} 
        \def\h{2.2} 
        \def\labelfar{0.3} 
        \def\angleA{130} 
        \def\angleB{50}  
        
        \draw[thick, fill=blue!5] (O) circle (\Re);
        \node[text=gray] at (0, -1) {\Large Earth};
        
        \coordinate (A) at (\angleA:\Re);
        \coordinate (B) at (\angleB:\Re);
        \coordinate (SA) at (\angleA:\Re+\h);
        \coordinate (SB) at (\angleB:\Re+\h);
        
        \coordinate (AP) at (\angleA:\Re+\labelfar);
        \coordinate (BP) at (\angleB:\Re+\labelfar);
        
        \draw[dashed, gray] (O) -- (A) node[midway, left=2pt] {$R_E$};
        \draw[dashed, gray] (O) -- (B) node[midway, right=2pt] {$R_E$};
        
        \draw[thick, red, <->] (\angleB:\Re) arc (\angleB:\angleA:\Re) node[midway, below=3pt] {$D_G$};
        
        \draw[ultra thick, cyan, ->] (A) -- (SA) node[midway, left] {$z_A \approx h$};
        
        \draw[ultra thick, cyan, ->] (B) -- (SB) node[midway, right] {$z_B \approx h$};
        
        \draw[ultra thick, orange, <->, dashed] (SA) -- (SB) node[midway, above] {$L_{\text{ISL}}$};
        
        \filldraw (O) circle (1.5pt) node[below] {$O$};
        
        \filldraw (A) circle (1.5pt);
        \filldraw (B) circle (1.5pt);
       
        \node[below left] at (AP) {Alice ($A$)};
        \node[below right] at (BP) {Bob ($B$)};
        
        \begin{scope}[shift={(SA)}, rotate=\angleA-90]
            \draw[fill=blue!40] (-0.6,-0.05) rectangle (-0.2,0.05); 
            \draw[fill=blue!40] (0.2,-0.05) rectangle (0.6,0.05);  
            \draw[fill=gray!20] (-0.2,-0.15) rectangle (0.2,0.15);  
            \node at (0, 0.4) {$S_A$};
        \end{scope}
        
        \begin{scope}[shift={(SB)}, rotate=\angleB-90]
            \draw[fill=blue!40] (-0.6,-0.05) rectangle (-0.2,0.05); 
            \draw[fill=blue!40] (0.2,-0.05) rectangle (0.6,0.05);  
            \draw[fill=gray!20] (-0.2,-0.15) rectangle (0.2,0.15);  
            \node at (0, 0.4) {$S_B$};
        \end{scope}
        
        \pic [draw, purple, "$\alpha$", angle radius=0.9cm, angle eccentricity=1.3] {angle = B--O--A};
    \end{tikzpicture}
}

\vspace*{8pt}

\caption{Geometry of the proposed dual-uplink dual-satellite architecture. By utilizing an Inter-Satellite Link ($L_{\text{ISL}}$), satellites $S_A$ and $S_B$ can be positioned at very small zenith angles (ideally $\theta \approx 0$) relative to their respective ground stations. This effectively relaxes the geometric constraints of single-node setups, minimizing atmospheric propagation distance such that $z \approx h$.}

\label{fig:DualLinkGeometry}

\end{figure}
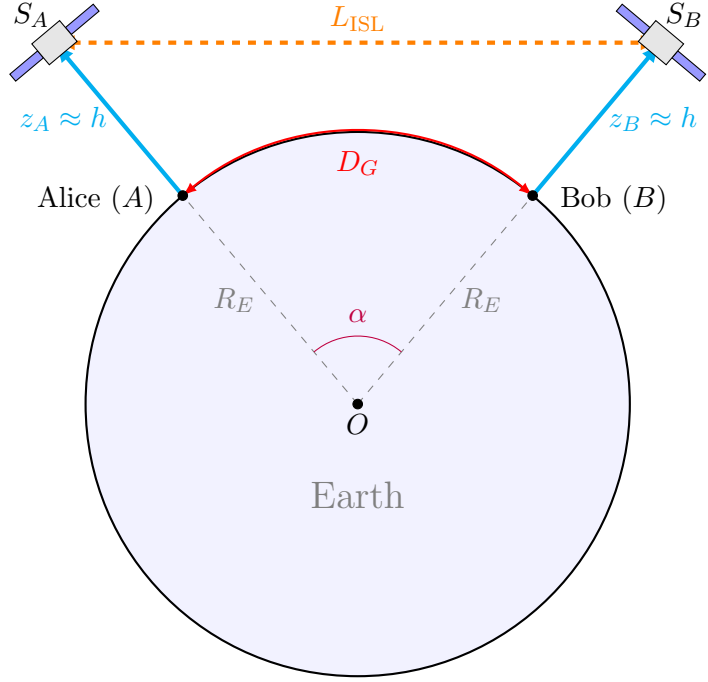

In a conventional single-satellite architecture, a single orbiting node must simultaneously establish optical links with both ground stations. This imposes a strict geometric dependence: if the satellite's position is optimized for station $A$ (e.g., minimizing its zenith angle), it inevitably increases the zenith angle and slant range for station $B$. The proposed dual-satellite architecture depicted in Fig. \ref{fig:DualLinkGeometry} fundamentally relaxes this geometric constraint by decoupling the two uplinks.

Let us define the geometric parameters of the system relative to the Earth's center $O$:

\begin{itemlist}
    \item $R_E$: The radius of the Earth (approximately $6371$ km).
    \item $h$: The orbital altitude of the satellites ($S_A$ and $S_B$).
    \item $D_G$: The great-circle distance separating the two ground stations, Alice ($A$) and Bob ($B$).
    \item $z_A, z_B$: The slant ranges (Euclidean propagation distances) for the uplinks.
    \item $L_{\text{ISL}}$: The straight-line propagation distance of the Inter-Satellite Link (ISL).
\end{itemlist}

As illustrated in Fig. \ref{fig:DualLinkGeometry}, the dual-satellite scheme allows each satellite to be positioned at a very small zenith angle relative to its respective ground station (ideally $\theta_A, \theta_B \approx 0$). Consequently, the uplink slant ranges are approximately equal to the satellite altitude $z_A \approx z_B \approx h$. This near-vertical configuration minimizes the optical path length through the atmosphere, drastically reducing turbulence-induced beam widening and atmospheric absorption. While this approximation ($z \approx h$) holds conceptually, our numerical channel model strictly computes the exact geometric slant range $z(h, \theta)$ to ensure absolute mathematical rigor in the evaluation of spatial and atmospheric losses.

The entanglement distribution is completed via the ISL connecting $S_A$ and $S_B$. Assuming the satellites are positioned at zenith relative to their respective ground stations, the central angle subtended by the two stations at the Earth's core is $\alpha = D_G / R_E$. Because the optical inter-satellite link signal propagates in a strict straight line through space, the exact propagation distance $L_{\text{ISL}}$ corresponds to the chord length between the two satellites at altitude $h$, given by:

\begin{equation}
\label{eq:isl_distance_exact}
    L_{\text{ISL}} = 2 (R_E + h) \sin\left( \frac{D_G}{2 R_E} \right)
\end{equation}

\subsection{Beam Deformation and Turbulence Effect}

As the optical beam propagates through the atmosphere, it undergoes spatial distortions that significantly reduce the photon collection probability at the receiver. These distortions are primarily classified into two phenomena \cite{alaluf2022,Bonato2009FeasibilityOS,Pirandola2020SatelliteQC,Srikara2024QuantumED}:

\begin{romanlist}[(ii)]
    \item {\bf Beam Widening:} The expansion of the beam's cross-sectional area. This is caused by two factors: the natural diffraction of light (which occurs even in a vacuum) and the scattering caused by atmospheric turbulence eddies \cite{andrews1993,born1999}.
    \item {\bf Beam Wandering:} The random fluctuation of the beam's centroid (center of intensity) relative to the receiver's optical axis. This is caused by large-scale turbulent eddies acting as refractive prisms that deflect the entire beam path \cite{fante1975,Pirandola2020SatelliteQC}.
\end{romanlist}

In an uplink configuration, the optical beam traverses the turbulent atmospheric layer immediately upon transmission. Small angular deviations induced by refractive index fluctuations near the ground are amplified by the long propagation distance $z$ to the satellite. Consequently, turbulence-induced beam wandering becomes the dominant loss mechanism \cite{Pirandola2020SatelliteQC,Srikara2024QuantumED}.

The effective long-term beam width $w_{up}$ at the satellite receiver is modeled by the superposition of vacuum diffraction and turbulence-induced broadening. The squared beam radius is given by \cite{Pirandola2020SatelliteQC,Bonato2009FeasibilityOS}:

\begin{equation} \label{eq:uplink_width}
w_{up}^2 = w_0^2 \left( 1 + \left(\frac{z\lambda}{\pi w_0^2}\right)^2 \right) + 2 \left(\frac{\lambda z}{\pi \kappa_0}\right)^2
\end{equation}

where:

\begin{itemlist}
    \item $w_0$ is the initial beam waist radius at the ground transmitter.
    \item $\lambda$ is the wavelength of the optical signal.
    \item $z$ is the slant range (propagation distance) of the uplink.
    \item $\kappa_0$ is the turbulence-dependent coherence length.
\end{itemlist}

The coherence length $\kappa_0$ encapsulates the strength of the atmospheric turbulence along the path and is highly sensitive to the zenith angle $\theta$. It is defined as:

\begin{equation} \label{eq:kappa0}
\kappa_0 = \left[ 1.46 \left(\frac{2\pi}{\lambda}\right)^2 \sec(\theta) \times C_w \right]^{-3/5}
\end{equation}

The term $\sec(\theta)$ accounts for the increased air mass at lower elevation angles. The factor $C_w$ represents the integrated refractive index structure constant profile ($C_n^2$) and typically takes values of $C_w \approx 3.28 \times 10^{-12} \, \text{m}^{1/3}$ during the day and $2.23 \times 10^{-12} \, \text{m}^{1/3}$ at night \cite{Pirandola2020SatelliteQC}.

The resulting average collection efficiency $\eta_{up}$, representing the probability that a photon enters the satellite's telescope aperture $R_{sat}$, is given by \cite{Pirandola2020SatelliteQC,Bonato2009FeasibilityOS,bourgoin2013}:

\begin{equation} \label{eq:uplink_efficiency}
\eta_{up} = 1 - \exp\left( \frac{-2R_{sat}^2}{w_{up}^2 + \sigma_{pe}^2} \right)
\end{equation}

where $\sigma_{pe}$ accounts for the pointing and tracking errors of the transmitter.

\subsubsection*{Inter-Satellite Link (ISL) and Total Geometric Losses}

In contrast to the Earth-to-space channel, the inter-satellite link (ISL) operates entirely in the vacuum of space. Consequently, the optical signal is completely immune to atmospheric turbulence. The beam propagating between the two satellites does not experience turbulence-induced wandering or scattering. Its expansion is governed exclusively by fundamental vacuum diffraction. 

Therefore, the squared beam radius for the ISL, $w_{ISL}^2$, over a propagation distance $L_{ISL}$, is simplified to the diffraction term \cite{Pirandola2020SatelliteQC}:

\begin{equation} \label{eq:isl_width}
w_{ISL}^2 = w_0^2 \left( 1 + \left(\frac{L_{ISL} \lambda}{\pi w_0^2}\right)^2 \right)
\end{equation}

Similarly, the collection efficiency for the inter-satellite link, $\eta_{ISL}$, follows the standard geometric formulation, utilizing the vacuum-broadened beam width $w_{ISL}$:

\begin{equation} \label{eq:isl_efficiency}
\eta_{ISL} = 1 - \exp\left( \frac{-2R_{sat}^2}{w_{ISL}^2 + \sigma_{pe}^2} \right)
\end{equation}

In the proposed dual-satellite architecture the total spatial/geometric collection efficiencies for the two entangled photons depend on their respective paths. The photon from ground station $A$ must survive both its vertical uplink to $S_A$ and the subsequent ISL to $S_B$. Conversely, the photon from ground station $B$ only needs to traverse its vertical uplink to $S_B$. The total spatial collection efficiencies for the Alice and Bob links are thus summarized as:

\begin{align} 
\eta_{A, geo} &= \eta_{up, A} \cdot \eta_{ISL} \label{eq:eta_a_geo} \\
\eta_{B, geo} &= \eta_{up, B} \label{eq:eta_b_geo}
\end{align}

\subsection{Stray Photon Noise}

In realistic satellite quantum links, multiple sources of background radiation contribute stray photons that must be carefully considered \cite{Bonato2009FeasibilityOS,pirandola2021limits,49,50,51,52}. During daytime operation, the dominant contributions arise from sunlight reflected by the Earth’s surface, whereas during nighttime, background photons originate primarily from reflected moonlight and the Earth’s intrinsic thermal emission \cite{Pirandola2020SatelliteQC,Bonato2009FeasibilityOS,50}. When these unwanted photons reach the satellite receiver simultaneously with the signal photons, they may introduce detection errors and degrade the overall measurement fidelity.

Let $r_{\mathrm{day}}$ and $r_{\mathrm{night}}$ denote the average arrival rates of stray photons at the photodetectors during daytime and nighttime conditions, respectively. Considering a detector time-gating window of duration $\tau = t_{\max} - t_{\min}$, the probability that $n$ background photons impinge on a single detector within this interval follows Poissonian statistics and is given by \cite{srikara2025supp}

\begin{equation}
P_{\mathrm{SP}}(n) = \frac{(r_{\mathrm{day/night}}\, \tau)^n e^{-r_{\mathrm{day/night}}\, \tau}}{n!}.
\label{eq:psp}
\end{equation}

\subsection{Atmospheric Attenuation}

The transmissivity of the atmospheric channel, denoted as $\eta_{atm}$, for a satellite link with zenith angle $\theta$, is given by \cite{Pirandola2020SatelliteQC,47,48}:

\begin{equation} \label{eq:attenuation}
\eta_{atm} = \exp\left[ -\alpha_0 \int_{0}^{z(h,\theta)} \exp\left( -\frac{h(y, \theta)}{\tilde{h}} \right) dy \right]
\end{equation}

where:

\begin{itemlist}
    \item $\alpha_0(\lambda)$: The extinction coefficient at sea level (equal to $5 \times 10^{-6} \, m^{-1}$).
    \item $z(h, \theta)$: The total slant range distance to the satellite.
    \item $\tilde{h}$: The atmospheric scale height (typically $\approx 6600$ m), representing the altitude at which atmospheric density drops by a factor of $1/e$.
    \item $h(y, \theta)$: The altitude of the beam at a distance $y$ along the path.
\end{itemlist}

This model captures the dominant effect of molecular absorption and scattering under clear-sky conditions.

\subsection{Mode Mismatch and Temporal Synchronization}

Successful Bell-state measurements (BSMs) require the two photons arriving from the ground stations to be indistinguishable in all degrees of freedom. In particular, temporal mismatch between the two wavepackets introduces distinguishability, reducing the quality of two-photon interference and consequently degrading the entanglement fidelity \cite{rohde2005,rohde2011}.

Let $\Delta t$ denote the relative arrival-time difference between the two photons. A time-gating window $[t_{\min}, t_{\max}]$ is employed at the detectors to ensure that only overlapping temporal components contribute to the measurement. While narrower gating improves indistinguishability, it also reduces the probability of successful detection, leading to a trade-off between fidelity and success probability.

Under Gaussian wavepacket assumptions and ideal channel conditions (i.e., neglecting other noise sources), the post-selected two-qubit state after the BSM is characterized by an interference term $\gamma$ and collection probabilities $P_{gw_1}$ and $P_{gw_2}$ corresponding to the two modes. The resulting fidelity with respect to the target Bell state is given by \cite{Srikara2024QuantumED}

\begin{equation}
F_{ic} = \frac{1}{2} + \frac{\gamma}{2 P_{gw_1} P_{gw_2}},
\label{eq:fidelity_mm_compact}
\end{equation}

where $\gamma$ quantifies the temporal overlap between the two wavepackets within the gating window, and $P_{gw_i}$ represents the probability that photon $i$ is detected within this window.

In the limit of perfect temporal overlap ($\gamma = P_{gw_1} P_{gw_2}$), the fidelity reaches unity. As the mismatch increases, the overlap term decreases and the fidelity approaches the classical bound of $0.5$.

\subsection{Overall Success Probability ($\eta_{tot}$)}

\begin{figure}[t]

\centerline{\includegraphics[width=4in]{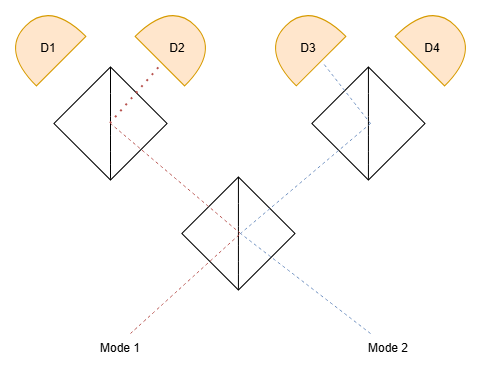}}

\vspace*{8pt}

\caption{Bell state measurement implementation inside the satellite.}

\label{fig:BSMSchema}

\end{figure}

We consider a linear-optical Bell-state measurement (BSM) implemented onboard the satellite (Fig.~\ref{fig:BSMSchema}). The setup consists of a beam splitter followed by polarization-resolving single-photon detectors, resulting in four detection channels $D_1, D_2, D_3,$ and $D_4$.

A detection outcome is described by the binary tuple 

\begin{equation}
d=(d_1,d_2,d_3,d_4), \quad d_i \in \{0,1\},
\end{equation}

where $d_i=1$ denotes a click event at detector $D_i$. Since the detectors cannot discriminate between signal and background photons, a click occurs whenever either a signal photon or a stray photon is present,

\begin{equation}
d_i = g_i \lor n_i ,
\end{equation}

with $g_i$ and $n_i$ denoting signal and noise events, respectively.

To elaborate, a successful projection onto the identifiable Bell states requires the two photons to emerge in different spatial modes at the beam splitter. Assuming detectors $D_1$ and $D_2$ monitor the horizontal and vertical polarization components of one spatial output mode, while $D_3$ and $D_4$ monitor the corresponding polarizations of the other spatial mode, valid signatures mandate exactly one click in each spatial mode. 

For example, a simultaneous detection in $D_1$ and $D_2$ (represented by the tuple $d=(1,1,0,0)$) indicates that both photons exited through the same port, which is an invalid signature. Conversely, a valid signature pattern would be a simultaneous detection in $D_1$ and $D_3$, represented by $d=(1,0,1,0)$. This indicates that the photons exited through different ports, successfully projecting the system onto a Bell state.

A successful Bell-state projection corresponds to specific valid detection patterns (or coincidence signatures), denoted as $M$. In other words, a detection pattern $M$ is simply a specific tuple $d$ (such as the $d=(1,0,1,0)$ example above) that satisfies the physical requirements for a successful BSM. The probability of observing a given signature $M$ is obtained by summing over all mutually exclusive signal--noise combinations that generate it,

\begin{equation}
P_M = \sum_{(g,n)\in \mathcal{C}_M} P_G(g)\, P_D(n),
\label{eq:success_prob_general}
\end{equation}

where $\mathcal{C}_M$ denotes the set of signal--noise pairs consistent with the detection pattern $M$, and $P_G$ and $P_D$ represent the signal and noise probability distributions, respectively \cite{Srikara2024QuantumED}.

For a symmetric optical configuration, four equivalent coincidence signatures correspond to successful Bell-state projections. The overall uplink efficiency is therefore given by

\begin{equation}
\eta_{tot} = \sum_{M \in \mathcal{S}} P_M = 4\, P_M^{(\mathrm{sym})},
\label{eq:eta_tot}
\end{equation}

where $\mathcal{S}$ is the set of valid coincidence patterns and $P_M^{(\mathrm{sym})}$ denotes the probability of any one representative signature under symmetry assumptions.

While all terms contributing to $P_M$ increase the raw detection rate, only the signal-only components preserve entanglement fidelity. Hybrid and noise-only contributions (e.g., stray photons triggering a valid pattern) lead to erroneous Bell-state projections and degrade the final state purity.

\subsection{Practical Fidelity ($F$)}

While the Overall Success Probability ($\eta_{tot}$) quantifies how often the satellite registers a valid detection signature, the {\it Fidelity} ($F$) quantifies the quality of the distributed entanglement for those successful events. In an ideal noise-free environment, the fidelity is limited only by the spatiotemporal indistinguishability of the photons. However, in a realistic uplink channel, background noise introduces ``false positive" signatures that severely degrade the purity of the final state.

We define the Practical Fidelity $F$ as a weighted average of the fidelity of  ``legitimate" events and ``illegitimate" events. Let $P_S$ be the conditional probability that a recorded valid signature (e.g. $M=(1,0,1,0)$) is caused by actual signal photons from the ground stations, rather than stray light or dark counts. This probability essentially represents the Signal-to-Noise Ratio (SNR) of the detection logic.

The final state distributed between Alice and Bob, $\rho_{final}$, is a mixture of the desired entangled state (created when signal photons interfere) and a maximally mixed state (created when noise triggers the detectors). The fidelity is given by \cite{Srikara2024QuantumED}:

\begin{equation} \label{eq:practical_fidelity}
F = P_S F_{ic} + (1 - P_S) F_{noise}
\end{equation}

\section{Dual-Uplink Dual-Satellite Architecture}\label{sec4}

In the following section, we present a detailed evaluation of the proposed dual-uplink dual-satellite architecture and provide a direct, quantitative comparison with the single-satellite baseline, utilizing the comprehensive performance results reported by Srikara {\it et al.} \cite{Srikara2024QuantumED}.

\subsection{The Protocol}

\begin{figure}[t]

\centerline{\includegraphics[width=4in]{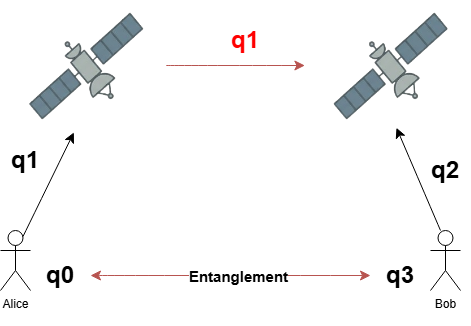}}

\vspace*{8pt}

\caption{Alice sends one qubit to the first satellite. The first satellite passively relays the qubit to the second satellite. Bob sends one qubit to the second satellite. The second satellite will perform the BSM to complete the entanglement swapping protocol.}

\label{fig:S2}

\end{figure}

The fundamental advantage of the proposed dual-satellite architecture lies in its network topology, which decouples the ground station separation distance ($D_G$) from the uplink transmission geometry. By utilizing two orbital nodes, the protocol allows each ground station to communicate with its local satellite at near-zenith elevation angles ($\theta \approx 0$), regardless of the total separation $D_G$. This vertical configuration ensures that the atmospheric path length remains minimal and effectively constant as the network scales. It turns out that the proposed protocol enables viable quantum entanglement distribution over intercontinental distances ($D_G > 1500$ km), a scale of operation that remains out of reach for single-satellite systems.

To bridge the spatial gap between the two orbital nodes, the architecture introduces an Inter-Satellite Link (ISL). Within our simulation framework, the ISL is strictly modeled as a free-space optical channel operating in vacuum. Because it is positioned entirely outside the Earth's atmosphere, turbulence-induced parameters, such as the coherence length ($\kappa_0$), are effectively removed from the ISL geometric loss calculations. Instead, the signal attenuation in the ISL is governed exclusively by deterministic free-space diffraction (fundamental beam widening) over the propagation distance $L_{\text{ISL}}$, alongside platform-induced pointing errors ($\sigma_{pe}$). 

Finally, regarding the optical relay within the satellites, we assume that mirror reflections do not inherently degrade the quality of photonic entanglement. Any potential fidelity degradation due to mirror recoil or polarization-dependent effects is considered negligible compared to the dominant spatial loss and geometric decoherence mechanisms present in satellite links \cite{ref17}.

\subsection{Model Assumptions}

\begin{itemlist}
    \item {\bf Polarization \& Doppler:} We neglect polarization errors, as they have been shown to be negligible ($<0.06\%$) \cite{Srikara2024QuantumED,Bonato2006InfluenceOS}, and we assume that Doppler shifts are compensated by precise wavelength calibration \cite{liu1999,rouzegar2019,Srikara2024QuantumED}.
     \item {\bf Nighttime Operation:} It must be emphasized that our analysis is exclusively focused on nighttime operation. Quantitative assessments of daytime quantum uplinks consistently indicate that such links are currently prohibitive due to the overwhelming solar background noise \cite{Srikara2024QuantumED,simmons2024}.
\end{itemlist}

\subsection{Fixed Simulation Parameters}

\begin{table}[h]
\centering
\tbl{System parameters used in the numerical simulations.\label{tab:simulation_parameters}}

{\begin{tabular}{@{}lcc@{}}
\toprule
\textbf{Parameter} & \textbf{Symbol} & \textbf{Value} \\
\colrule
Satellite aperture diameter & $D$ & $150$ cm \\
Operating wavelength & $\lambda$ & $800$ nm \\
Detector efficiency & $\eta_{det}$ & $0.6$ \\
Total measurement efficiency & $\eta_{meas}$ & $0.25$ \\
PBS transmissivity & $\eta_{pbs}$ & $0.91$ \\
BSM efficiency factor & -- & $0.5$ \\
Earth's surface temperature & $T$ & $300$ K \\
Clock sync precision & $\Delta t_{sync}$ & $5$ ns \\
Beam Waist & $w_0$ & $13$ cm\\
Zenith Angle & $\theta$ & $0.1$ rad\\
\botrule
\end{tabular}}

\end{table}

The simulation parameters are summarized in Table \ref{tab:simulation_parameters}. All parameters correspond to values reported in recent experimental or feasibility studies and do not rely on speculative technological advances \cite{Pirandola2020SatelliteQC,Bonato2009FeasibilityOS,srikara2025supp,ref60}.

\section{Numerical Results} \label{sec5}

In this section, we present the numerical results derived from the simulations of the dual-satellite architecture. The analysis focuses on quantifying both the achievable transmission rate and the quality of the distributed entanglement. For all subsequent simulations, we fix the photon pulse width at $\sigma_t = 10$ ns and the gating window at $t_{gate} = 40$ ns.

\subsection{Performance at Regional Scales ($D_G \le 1500$ km)}

\begin{figure}[t]

\centerline{\includegraphics[width=4in]{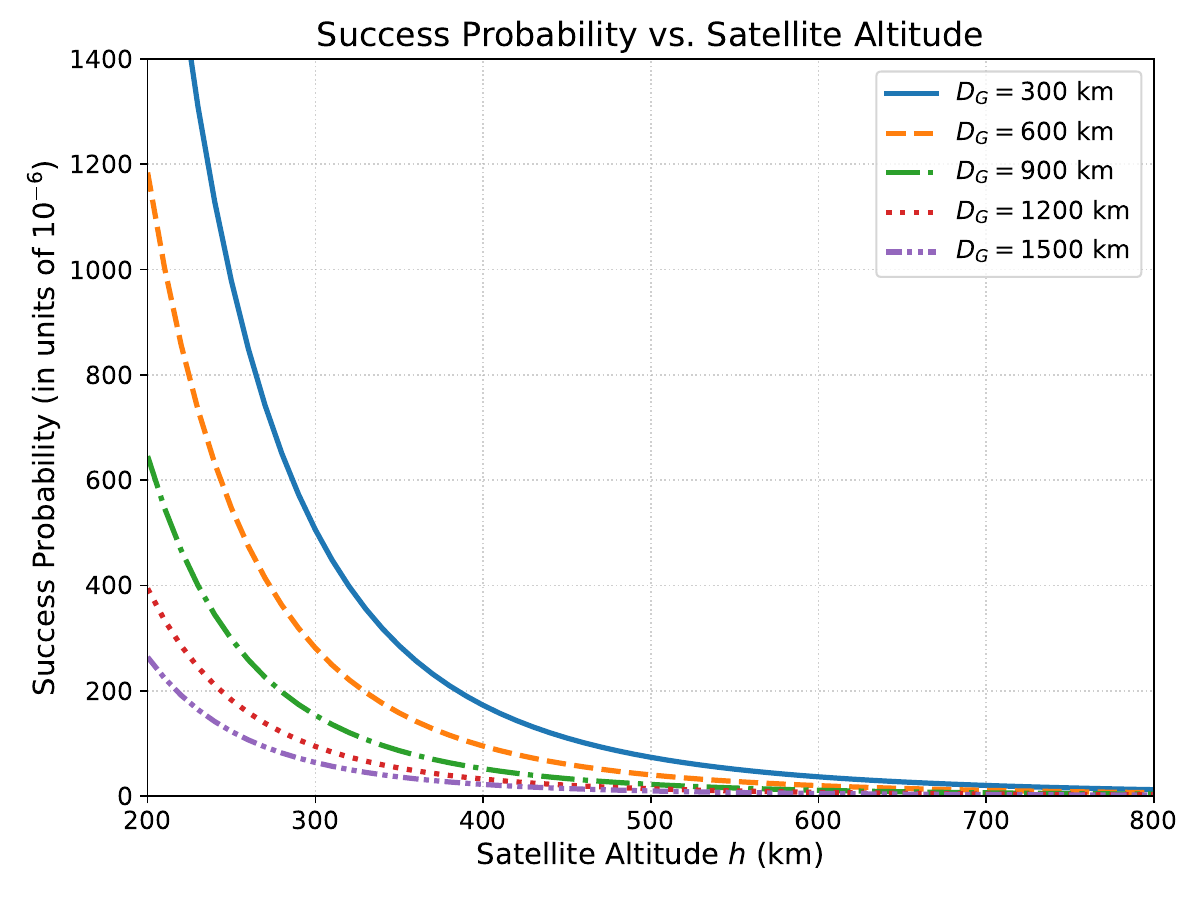}}

\vspace*{8pt}

\caption{Success probability as a function of satellite altitude for different ground station distances $D_G$.}

\label{fig:SuccDist100_1000}

\end{figure}

\begin{figure}[t]

\centerline{\includegraphics[width=4in]{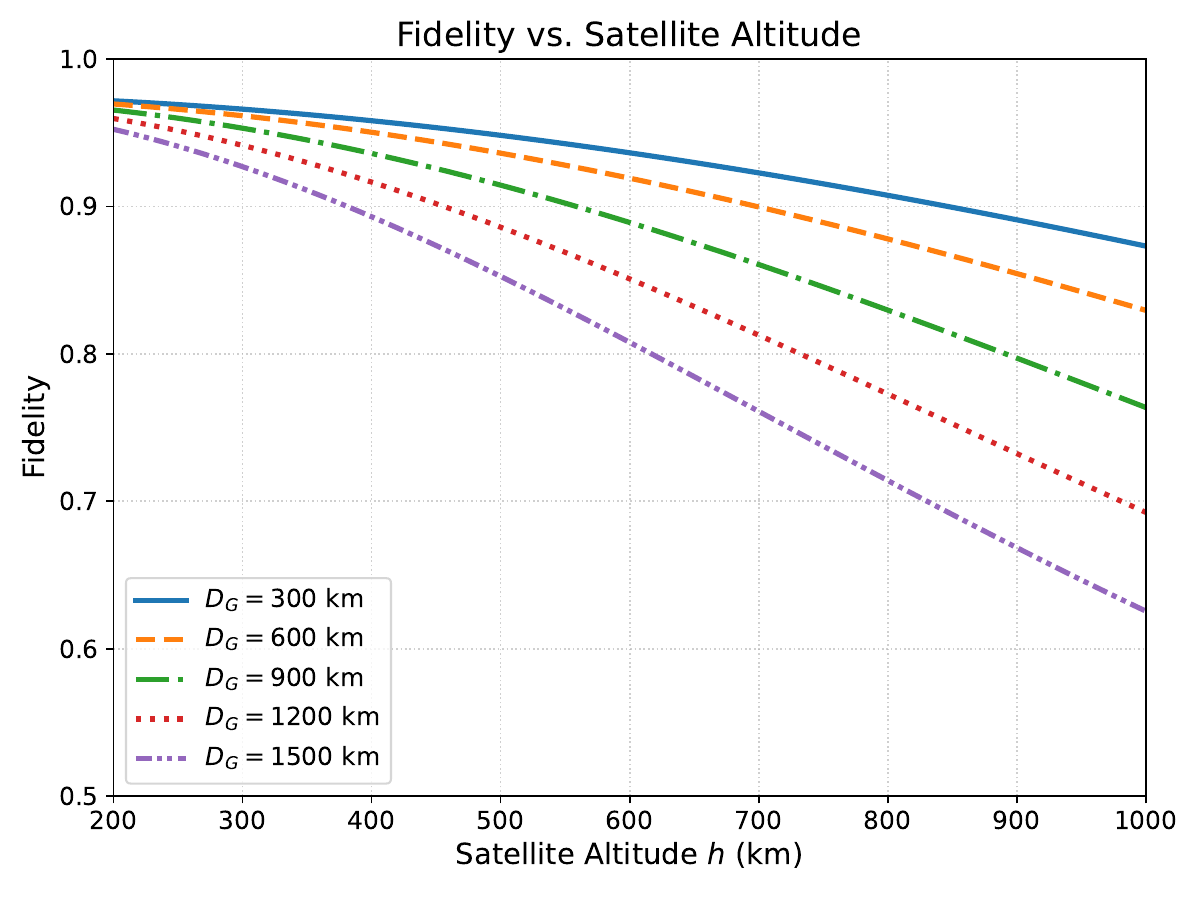}}

\vspace*{8pt}

\caption{Fidelity as a function of satellite altitude for different ground station distances $D_G$.}

\label{fig:FidDist100_1000}

\end{figure}

We begin by examining the system performance at shorter inter-ground station distances to provide a direct comparison between the single-satellite and dual-satellite architectures. The results for various ground station distances $D_G$ are illustrated in Fig. \ref{fig:SuccDist100_1000} and \ref{fig:FidDist100_1000}.

\begin{figure}[t]

\centerline{\includegraphics[width=4in]{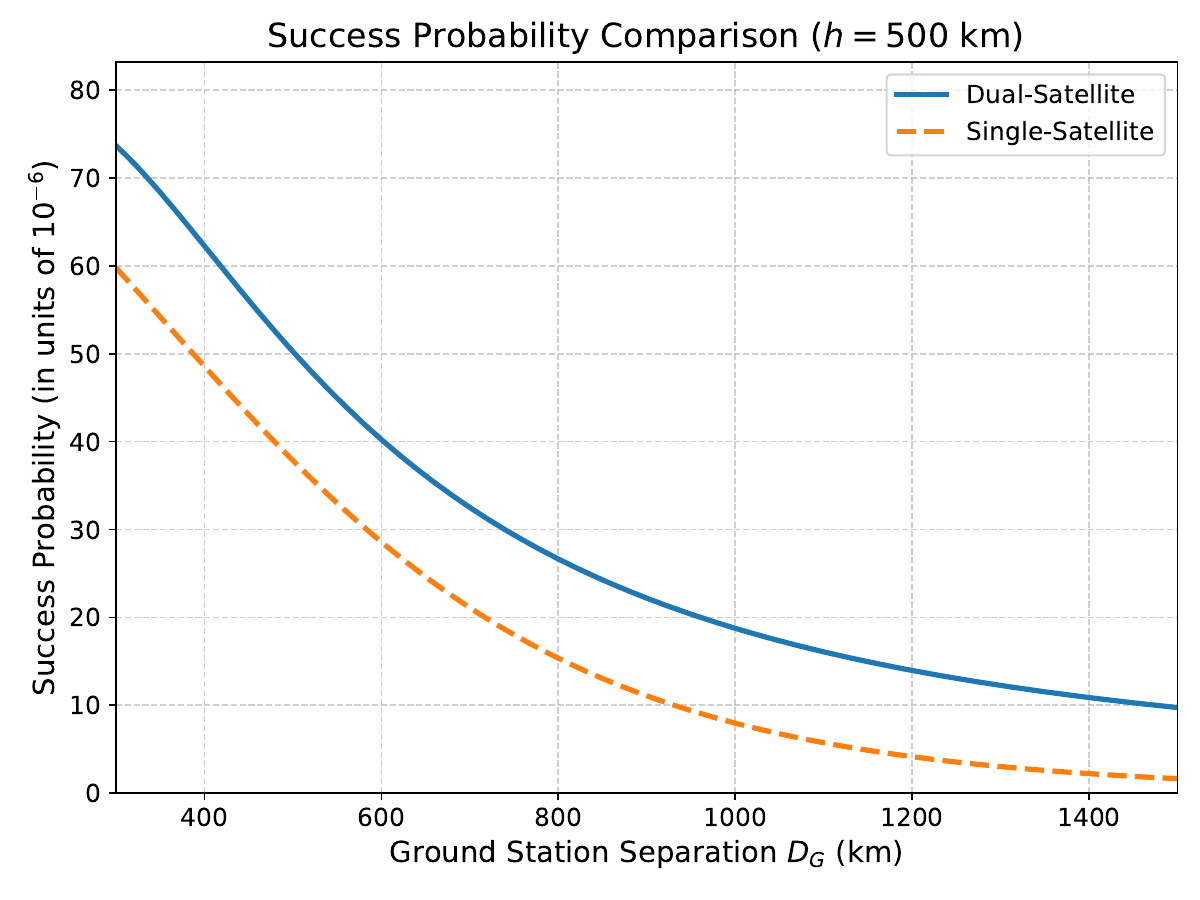}}

\vspace*{8pt}

\caption{Comparison of the single and dual-satellite architectures' success probability as a function of ground station distance for $h=500$ km.}

\label{fig:Success1500Compare_1and2}

\end{figure}

\begin{figure}[t]

\centerline{\includegraphics[width=4in]{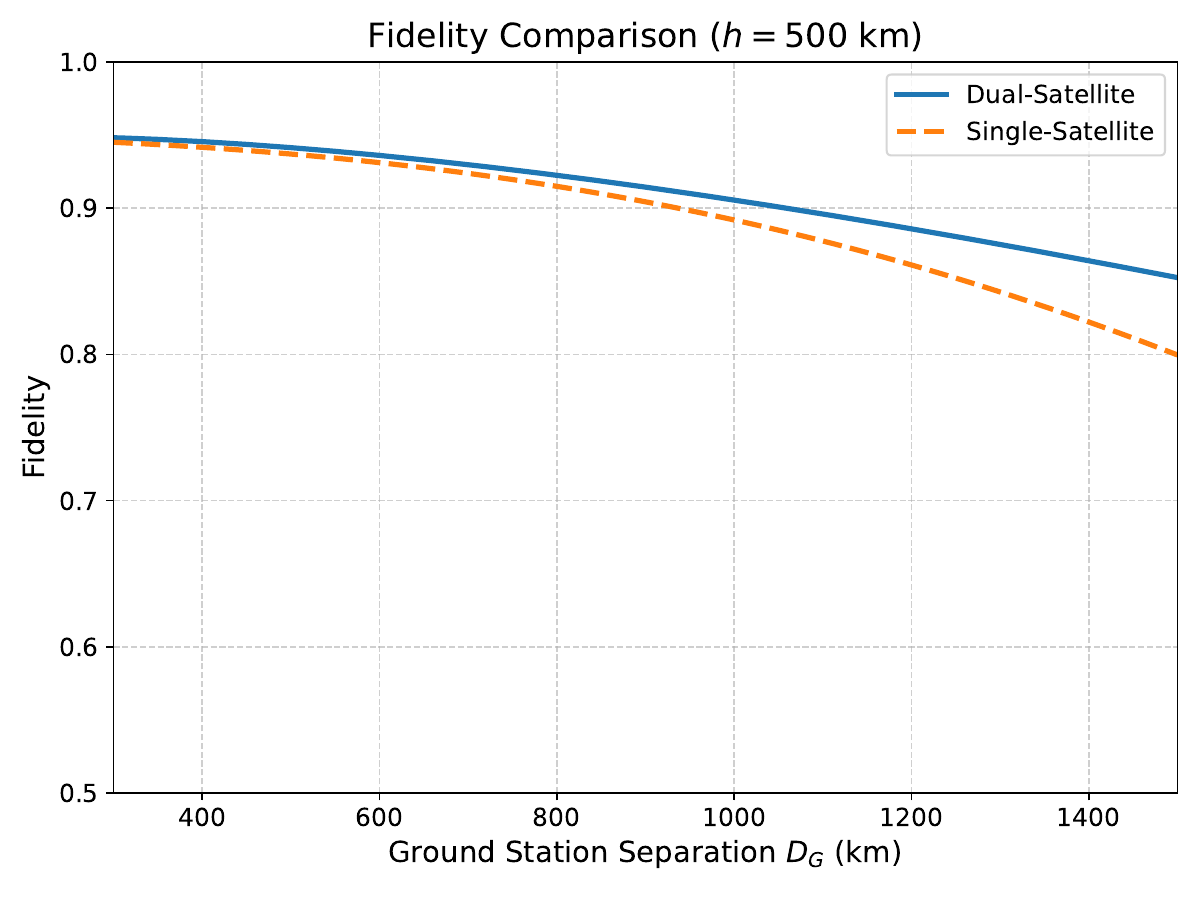}}

\vspace*{8pt}

\caption{Comparison of the single and  dual-satellite architectures' fidelity as a function of ground station distance for $h=500$ km.}

\label{fig:FidCompare1_2}

\end{figure}

The comparison of the single and dual-satellite architectures as a function of the ground station distance $D_G$ for a satellite altitude of $h=500$ km is illustrated in Fig. \ref{fig:Success1500Compare_1and2} and \ref{fig:FidCompare1_2}.

\vspace*{12pt}

\noindent

{\bf Overall Success Probability}

As observed in Fig. \ref{fig:SuccDist100_1000}, the success probability exhibits an inverse relationship with the satellite altitude $h$. Lower orbits (e.g. $h=500$ km) yield higher detection rates due to the minimized path length of the vertical uplinks.

Crucially, the dual-satellite architecture demonstrates a significant performance enhancement compared to the single-satellite baseline. For a specific case study at $h=500$ km and $D_G=500$ km (with identical optical parameters), the single-satellite architecture yields an overall success probability of approximately $3.34 \times 10^{-5}$, whereas the dual-satellite architecture achieves $5 \times 10^{-5}$. This improvement arises from the reduced atmospheric path length enabled by near-zenith uplinks.

\vspace*{12pt}

\noindent

{\bf Fidelity}

As observed in Fig. \ref{fig:FidDist100_1000}, for $h=500$ km, the fidelity remains remarkably robust, initiating at $F \approx 0.95$ and maintaining values above $0.90$ for $D_G \le$ $900$ km. In contrast, at $h=1000$ km, increased diffraction losses significantly weaken the signal, thereby reducing the signal-to-noise ratio at the detectors, causing the fidelity to drop toward $0.75$. 

The fidelity results at shorter distances for the dual-satellite topology are comparable to those of the single-satellite architecture. Revisiting the case study at $h=500$ km and $D_G=500$ km, the single-satellite architecture yields a fidelity of $\approx 0.937$, while the dual-satellite topology yields a fidelity of  $\approx 0.94$.

\subsection{Performance at Intercontinental Scales ($D_G \ge 1500$ km)}

\begin{figure}[t]

\centerline{\includegraphics[width=4in]{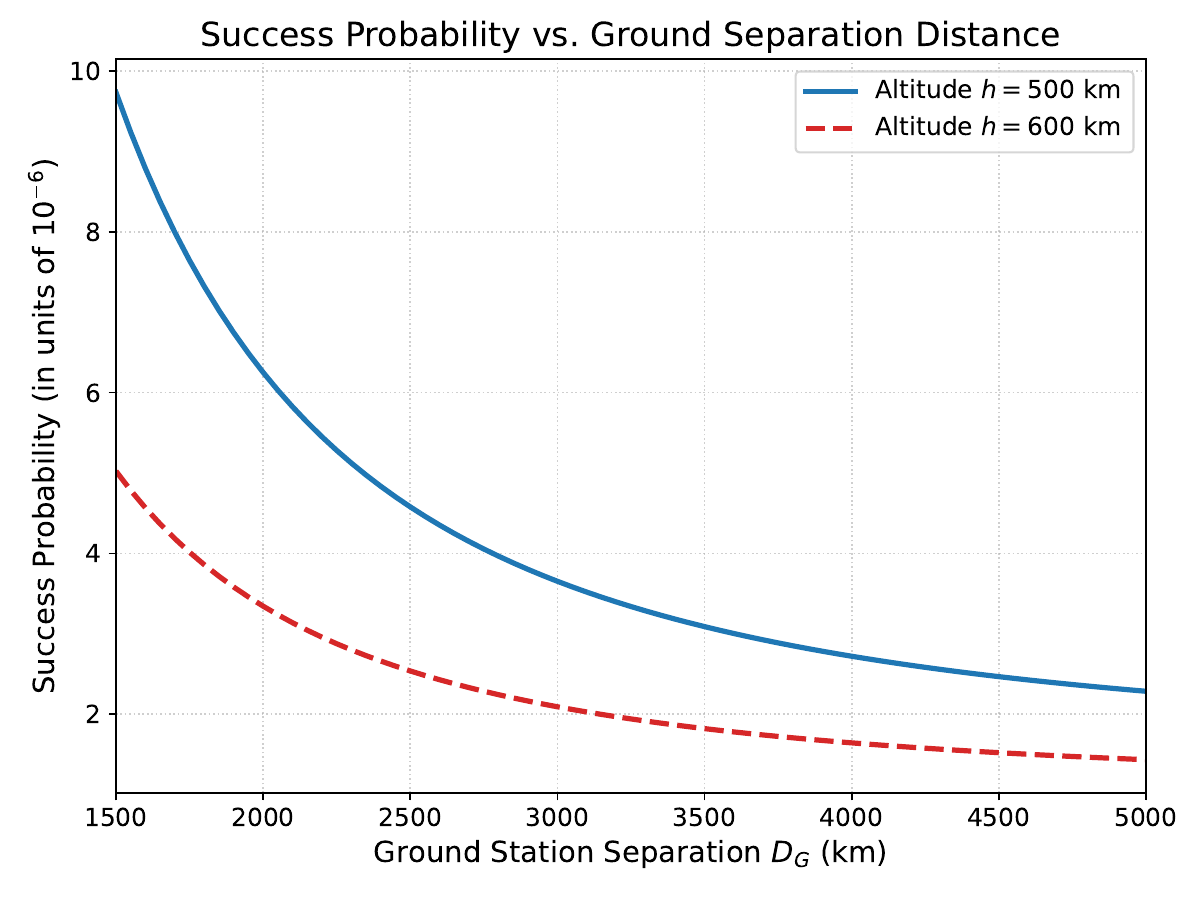}}

\vspace*{8pt}

\caption{Success probability as a function of ground distance for different satellite altitudes $h$.}

\label{fig:SuccDist1500_5000}

\end{figure}

\begin{figure}[t]

\centerline{\includegraphics[width=4in]{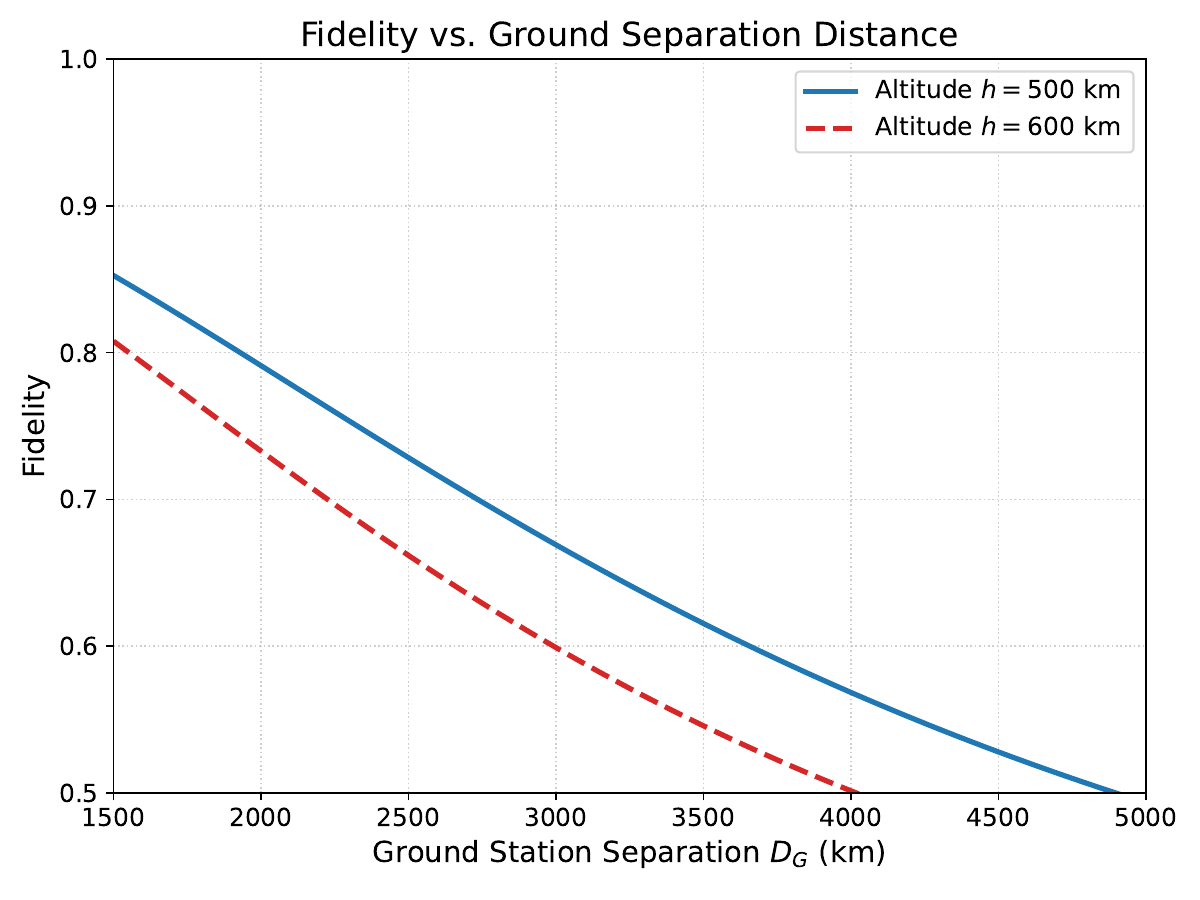}}

\vspace*{8pt}

\caption{Fidelity as a function of ground distance for different satellite altitudes $h$.}

\label{fig:FidDist1500_5000}

\end{figure}

Subsequently, we extend the analysis to greater ground distances, showcasing the fundamental advantage of the dual-node topology: its ability to bridge distances physically inaccessible to a single LEO node. The results are presented in Fig. \ref{fig:SuccDist1500_5000} and \ref{fig:FidDist1500_5000}.

\vspace*{12pt}

\noindent

{\bf Overall Success Probability}

The simulation results, observed in Fig. \ref{fig:SuccDist1500_5000}, indicate that operational success probabilities in the order of $10^{-6}$ are achievable for satellite altitudes of 500-600 km, even at distances approaching 5000 km. However, it should be noted that while the signal remains detectable at these ranges, the choice of altitude involves a trade-off: lower altitudes ($h=500$ km) maximize the signal strength but impose a stricter geometric horizon limit ($\sim 5000$ km), while slightly higher orbits ($h=600$ km) provide a safer geometric margin for intercontinental links at the cost of a marginal reduction in received flux.

\vspace*{12pt}

\noindent

{\bf Fidelity}

The results in Fig. \ref{fig:FidDist1500_5000}, show that as the ground distance extends into the intercontinental regime, the structural advantage of the dual-node architecture becomes evident, albeit facing significant challenges at extreme ranges. The primary factor limiting fidelity at these ranges is the reduction in the SNR. As the distance between the two satellites increases, the beam divergence causes the spot size to exceed the receiver's aperture, resulting in a signal that gradually approaches the noise floor of the detectors (dark counts). 

Specifically, for a satellite altitude of $h=600$ km, the fidelity stands at $F \approx 0.80$ at $1500$ km but degrades to $F \approx 0.44$ at $5000$ km. Although the fidelity at $5000$ km falls below the classical limit for secure communication, the dual-satellite architecture still outperforms the single-satellite configuration at intermediate intercontinental distances (e.g. at $2000$ km, $F_{dual} \approx 0.733$ vs $F_{single} \approx 0.68$), while also maintaining significantly higher success probabilities. This confirms that the replacement of stochastic atmospheric losses with deterministic vacuum diffraction provides a more ``graceful" degradation of the quantum state, extending the operational reach of the network.

\subsection{Operational limits}

Taking into account the results presented in the previous section, the dual-satellite architecture remains operational for distances up to around 2000 km. The fidelity at this point drops to 0.733, which approaches the theoretical lower bound for secure key distillation. Note that, although the success probability at larger distances remains at operational values, the low fidelity indicates that these detection events are primarily attributed to noise and dark counts rather than to the actual signal.

However, this is still a major step forward compared to the single-satellite architecture and serves as the base for further improvements.

\subsection{Advantages}

\begin{itemlist}
    \item {\bf Atmospheric Decoupling:} The most significant advantage is the decoupling of the total communication distance from atmospheric losses. Since the ground-to-satellite uplinks remain near-vertical ($\theta = 0.1$ rad), the atmospheric attenuation and turbulence-induced noise remain minimal and constant, regardless of whether the ground stations are separated by 100 km or 5000 km.
    \item {\bf Intercontinental Operational Range:} This architecture enables viable quantum links over intercontinental distances ($D_G > 2000$ km). It effectively bypasses the geometric horizon limits and the exponential signal collapse that plagues single-satellite configurations at low elevation angles.
    \item {\bf Deterministic Loss Profile in Vacuum:} Unlike the stochastic nature of atmospheric turbulence in the single-satellite architecture, the ISL segment in the dual-satellite architecture  operates in a vacuum. The losses are purely deterministic, governed by Gaussian beam diffraction ($\propto 1/L^2$), allowing for more predictable system performance and stable Signal-to-Noise Ratios (SNR).
    \item {\bf Superior Transmission Rates:} Due to the reduced total path loss at long distances, the dual-satellite architecture maintains a success probability in the order of $10^{-6}$, which is nearly an order of magnitude higher than the $10^{-7}$ range observed in the single-satellite architecture for the same distances.
\end{itemlist}

\subsection{Disadvantages}

\begin{itemlist}
    \item {\bf Hardware and Structural Complexity:} The dual-satellite architecture requires double the number of orbital nodes and significantly more complex payloads. Each satellite must be equipped with multiple optical terminals to manage simultaneous ground-to-space and space-to-space links.
    \item {\bf Stringent Temporal Synchronization:} Achieving the necessary temporal mode-matching for Bell State Measurements (BSM) is highly challenging. The assumed 5 ns jitter requires stringent nanosecond-level precision in clock synchronization between two independent, high-velocity orbital nodes. Such requirements are challenging, but not fundamentally prohibitive, with current or near-term technology.
    \item {\bf Orbital Dynamics and Point-Ahead Angle (PAA):} Targeting a laser beam between two orbiting satellites requires microradian pointing precision. A sophisticated Point-Ahead Angle mechanism is mandatory to compensate for the displacement of the receiver during the photon's time-of-flight.
    \item {\bf Vacuum Diffraction Limits:} While vacuum propagation avoids turbulence, beam divergence (widening) remains a limiting factor.
\end{itemlist}

\section{Conclusion} \label{sec6}

This study investigated a dual-uplink dual-satellite architecture designed to overcome the fundamental geometric limitations of the single-satellite uplink entanglement distribution. By comparing this approach against the established single-satellite model under a unified simulation framework, we were able to isolate the effects of geometry and atmospheric turbulence on link performance.

Our results highlight a fundamental bottleneck in the single-satellite approach: the requirement for simultaneous visibility creates a ``hard ceiling" on scalability, making it ineffective beyond regional distances. The transition to a dual-satellite architecture effectively breaks this geometric dependency by decoupling the atmospheric uplink path from the total ground station separation. We demonstrated that with the added complexity of even one inter-satellite link, this configuration offers a distinct advantage in success probability and allows the network to bridge distances that are physically impossible for a single node.

However, this advantage comes at the cost of increased synchronization demands and orbital complexity. These challenges do not invalidate the proposed architecture but instead define the primary engineering constraints that must be addressed. 

Although the dual-satellite model is not a complete solution to all channel loss challenges, our findings suggest that it is a necessary evolutionary step. Moving from single-hop to multi-hop satellite networks appears to be the only viable path for realizing a truly global quantum internet.

Future work should focus on optimized orbital configurations, advanced inter-satellite synchronization schemes, and the extension of the proposed model to multi-hop satellite constellations.

\end{document}